\documentclass[prb,floats,aps,twocolumn]{revtex4}
\usepackage{graphicx}
\usepackage{amsmath}
\newcommand{\rr}{{\bf r}}
\newcommand{\qq}{{\bf q}}
\newcommand{\Q}{{\bf Q}}
\newcommand{\RR}{{\bf R}}
\newcommand{\kk}{{\bf k}}
\newcommand{\BSCCO}{{Bi$_2$Sr$_2$CaCu$_2$O$_8$}}

\newcommand{\sigmab}{\overline{\sigma}}
\begin{document}
\title{Local density of states of a
$d$-wave superconductor with inhomogeneous antiferromagnetic correlations}
\author{W. A. Atkinson} \affiliation{Trent University} \date{\today}
\begin{abstract}
The tunneling spectrum of an inhomogeneously doped extended Hubbard
model is calculated at the mean field level.  Self-consistent
solutions admit both superconducting and antiferromagnetic order,
which coexist inhomogeneously because of spatial randomness in the
doping. The calculations find that, as a function of doping, there is
a continuous cross over from a disordered ``pinned smectic'' state to a
relatively homogeneous $d$-wave state with pockets of
antiferromagnetic order.  The density of states has a robust $d$-wave
gap, and increasing antiferromagnetic correlations lead to a
suppression of the coherence peaks.  The spectra of isolated nanoscale
antiferromagnetic domains are studied in detail, and are found to be
very different from those of macroscopic antiferromagnets. Although no
single set of model parameters reproduces all details of the
experimental spectrum in \BSCCO, many features, notably the collapse
of the coherence peaks and the occurence of a low-energy shoulder in
the local spectrum, occur naturally in these calculations.
\end{abstract}

\maketitle

\section{Introduction}
Nanoscale inhomogeneities have been widely observed in the high
temperature superconductor (HTS) \BSCCO (BSCCO), primarily through
scanning tunneling microscopy (STM) experiments performed in the
superconducting
state.\cite{cren,pan,Kapitulnik1,lang,Kapitulnik2,Kapitulnik3,McElroy2}
At present, the origins of the inhomogeneity are not well understood,
although it is very plausible that they are directly correlated with
variations in the local doping concentration.  In BSCCO, the hole
concentration is controlled by the addition of interstitial oxygen
atoms which appear to reside $\approx 5$ \AA above the conducting
CuO$_2$ layers.  Because of the short distance,
large spatial fluctuations of the Coulomb potential are expected in
the CuO$_2$ layers, especially in underdoped samples where screening
is poor.  In STM experiments (which do not measure local doping
directly), nanoscale inhomogeneities are manifested most strongly in
the magnitude $\Delta_T$ of the superconducting gap in the tunneling
spectrum. Interestingly, regions with small $\Delta_T$ exhibit large
coherence peaks at the gap edge, while the coherence peaks are
essentially missing in regions with $\Delta_T > 65$ meV.  These latter
regions are assumed to represent an underdoped ``pseudogap'' phase
which may be quite distinct from the small-gap ``superconducting''
(SC) regions.  Though speculative, this labelling is supported by the
fact that the ``pseudogap'' regions occupy a large fraction of the
strongly underdoped samples, and relatively little of the optimally
doped samples.\cite{McElroy2}

The simplest model of the inhomogeneities is that the pairing energy
depends strongly on doping,\cite{zhangrice,engelbrecht,randeria} with
the ``pseudogap'' domains corresponding to regions of large pairing
energy which are nonsuperconducting because of phase fluctuations.  It
is also possible that secondary phases may coexist with the SC state,
forming spontaneously, or perhaps being nucleated in hole-poor
domains.  Theoretical calculations from the early days of
high-temperature superconductivity\cite{poilblanc,zaanen,schultz,kato}
suggest the possibility of self-organized stripe formation, and in
some materials (notably
La$_{2-x-y}$Nd$_y$Sr$_x$CuO$_4$\cite{stripeexpt}) there is solid
evidence for stripes, though it has generally been hard to
substantiate in other materials.  A large variety of other
competing or coexisting phases have been discussed since the
discovery of HTS, and the list includes
charge density wave (CDW),\cite{CDW,checkerboard} spin
Peierls,\cite{marston,Peierls} antiferromagnetic or spin density
wave,\cite{sczhang,polkovnikov,franz,alvarez} pair density
wave,\cite{checkerboard,ChenZhang}, staggered flux\cite{palee},
and orbital current\cite{marston,schulz,varma,chakravarty} phases.

Experimentally, optimally-doped BSCCO\cite{Hoffman1,McElroy,DHLee}
appears to support a fairly straightforward $d$-wave BCS picture of
superconductivity rather well,
\cite{DHLee,ZAH,Tingfourier,capriotti,franzftdos} although similar
experiments\cite{Kapitulnik2,Kapitulnik3} have been interpreted in
terms of commensurate stripe formation with a periodicity of $\approx
4a_0$ where $a_0\approx 5$ \AA is the lattice constant.  At lower
doping, other recent work\cite{McElroy2} finds weak ``checkerboard''
charge modulations with a periodicity close to that measured in
Refs.~[\onlinecite{Kapitulnik2,Kapitulnik3}], with the weight of the
modulations being reduced as the doping increases.  However, the
situation is not transparent since the modulations are only seen at
energies larger than the gap edge (contrary to what one might expect
in a stripe scenario), are only seen in the ``pseudogap'' regions, and
the modulation wavelength is comparable to the typical size of the
``pseudogap'' domains.
%
There is a further ambiguity in determining what, if any, ordering is
present: many ordered phases (eg.\ antiferromagnetism) which may be
relevant to BSCCO couple weakly to the local charge density and are
not easily identified in STM experiments.  Thus, it is not clear
whether the weak charge modulations seen in experiments are the
dominant ordering, or whether they are secondary manifestations of
some hidden order.  For these reasons, it may be difficult to detect
and study coexisting order based on spatial modulations of the 
local density of states (LDOS)
alone.  

The goal of the present work is to look for signatures of
inhomogeneously coexisting order in the energy dependence of the local
spectrum.  For definiteness, I adopt a model in which
antiferromagnetic (AF) correlations compete with SC order.  From a
calculational perspective, this is the simplest and least ambiguous
choice, although other order parameters---particularly CDW order
(including checkerboard order)---are also potentially relevant to the
STM experiments cited above.  Many of the results of this paper will
actually apply broadly to other forms of competing order, and I will
try distinguish these results from those which are specific to
antiferromagnetism.  Having said this, I want to remark that
antiferromagnetism is a natural choice to make given the proximity of
the AF and SC phases in the HTS phase diagram, and that other authors
have studied similar models.\cite{zaanen,sczhang,polkovnikov,alvarez}
In addition, there is mounting evidence that glassy (short-ranged)
quasistatic AF correlations are significant in underdoped
HTS,\cite{julien,sidis,mook,sanna,lake,panagopoulos} including
BSCCO,\cite{panagopoulos}  and it is important to understand how
these correlations are manifested in the LDOS.

The paper is organized as follows: In Sec.~\ref{sec2a}, I introduce
the model, and perform calculations for a finite-sized,
inhomogeneously doped $d$-wave superconductor with competing AF and SC
order.  Short range AF order arises naturally in the current work
because the system is doped inhomogeneously by charged out-of-plane
donors and AF moments form preferentially in underdoped regions.  At
low doping levels, I find that the self-consistently determined
electronic state resembles a ``pinned smectic" in which
superconductivity is pronounced along domain walls of the AF
background.  At higher doping, there is a crossover to a fairly
homogeneous $d$-wave SC state with occasional pockets of AF order.  In
all cases, there is a well defined $d$-wave gap in the spectrum.
Since the spectral energy resolution suffers from finite-size effects,
I discuss the LDOS in the context of a single underdoped pocket
embedded in a homogeneous $d$-wave superconductor in Sec.\ref{sec2b}.
Several spectral features measured in [\onlinecite{McElroy2}], notably
the suppression of coherence peaks, the appearance of shoulders in the
spectrum, and the homogeneity of the low energy spectrum, can be
understood in these calculations, although no single parameter set
reproduces simultaneously all the experimentally measured spectral
features.  One of the most important conclusions of this section is
that, because of the nonlocality of quasiparticles, the local spectrum
of an AF pocket resembles neither that of macroscopic antiferromagnets
or superconductors (nor is it an average of the two): The introduction
of inhomogeneity on the nanometer length scales leads to a
qualitatively new spectrum.  This is a significant finding since one
of the main arguments against coexisting secondary phases is that,
apart from the special case of a nested Fermi surface, any {\em
macroscopic} ordering which is commensurate with the lattice has a
spectrum which is not particle-hole symmetric, in contradiction with
experiments.  I find, however, that inhomogeneous ordering on
nanometer length scales {\em may}, in fact, yield a particle-hole
symmetric spectrum.  These calculations are interpreted in terms of a
three-band model of homogenously coexisting SC and AF order in
Sec.~\ref{sec2c}. The issue of how charge modulations arise in this
model is discussed in Sec.~\ref{sec2d}.  Conclusions are presented in
Sec.~\ref{sec3}.

\section{Calculations and Results}
\label{sec2}
\subsection{Inhomogeneously doped superconductor}
\label{sec2a}
The basic Hamiltonian is the Hubbard model with a long-range Coulomb
interaction and SC pairing interaction:
\begin{eqnarray}
H &=& \sum_{i,j,\sigma} t_{ij} c^\dagger_{i\sigma}c_{j\sigma} 
-Z \sum_{i,\RR}
V(\rr_i -\RR)\hat n_i \nonumber \\
&& + \sum_i U \hat n_{i\uparrow} \hat n_{i\downarrow} 
+\frac 12 \sum_{i\neq j} V(\rr_i-\rr_j) \hat n_i \hat n_j
\nonumber \\ &&
+\sum_{ij} \Delta_{ij} (c^\dagger_{i\uparrow}c^\dagger_{j\downarrow}
+c_{i\downarrow}c_{j\uparrow})
\label{HU}
\end{eqnarray}
where $c_{j\sigma}$ is the spin-$\sigma$ annihilation operator at site
$j$, $\hat n_{i\sigma}$ and $\hat n_i$ are the spin-resolved and
total charge density operators at site $i$, and $\rr_i$ is the position
of the $i^{\mathit{th}}$ site.  I use a third-nearest
neighbor conduction band with parameters $t_0,$\ldots$t_3$ describing
the on-site potential, nearest, next-nearest, and third-nearest
neighbor hopping amplitudes.  Throughout this work, all energies are
given in units of $|t_1|$ which (for reference) is $\sim O(100)$
meV.  I take $\{t_1,t_2,t_3 \} = \{-1,0.25,-0.1 \}$ and adjust $t_0$
to give the desired filling.  The long range Coulomb interaction is
$V(\rr) = (e^2/\epsilon a_0)|\rr|^{-1}$ where $\rr$ is measured in
units of the lattice constant $a_0$, $e^2/\epsilon a_0 = 1$, and the
on-site interaction is absorbed into the Hubbard $U$ term: $V(0) =
U/2$.
The impurities are located at positions $\RR$, which sit a distance
$d_z = 1.5a_0$ above randomly chosen lattice sites.  The final term in
the Hamiltonian is added as an ansatz to describe SC order
arising from spin-interactions between neighboring sites.  The local
bond order parameter $\Delta_{ij} = -\frac {J}{2} \langle
c_{j\downarrow} c_{i\uparrow} + c_{j\uparrow} c_{i\downarrow}\rangle$
is determined self-consistently for nearest neighbor sites $i$ and
$j$.  The Coulomb interaction is treated in the Hartree approximation,
and the effective mean-field Hamiltonian can be diagonalized
numerically to extract eigenstate wavefunctions and the corresponding
eigenenergies.  The fields $\Delta_{ij}$ and $n_{i\sigma}$ are
iterated to self-consistency on small lattices with between $20\times
20$ and $40\times 40$ sites.  The calculations are unconstrained, and
are seeded with a finite antiferromagnetic moment.  In order to
improve convergence, which is problematic when magnetic moments form,
a combination of Thomas-Fermi and Pulay method charge-mixing is used
at each iterative step.\cite{pulay}

\begin{figure}
\includegraphics[width=\columnwidth]{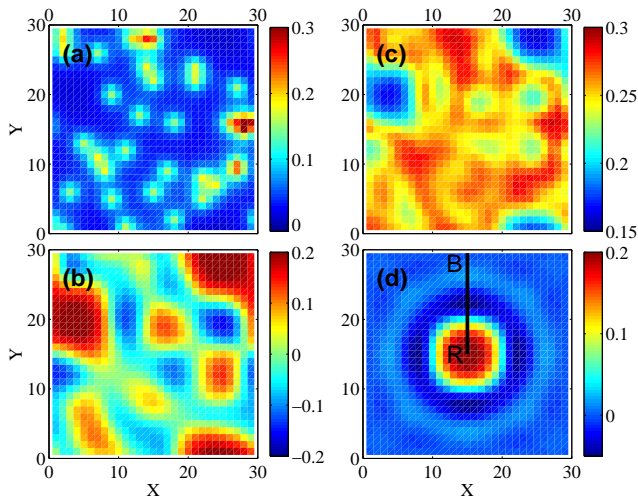}
\caption{Self-consistent solutions of the model.  (a) Charge density,
(b) staggered magnetization, and (c) d-wave gap are shown for a
$30\times 30$ lattice with 35 donor impurities of charge $Z=-2e$.  The
model parameters are $U=3.2$ and $J=1.5$.  The corresponding hole
doping level is $n_h \approx 0.07$.  The staggered magnetization for a
single underdoped disk of radius $4a_0$ is shown in (d).  Notice that
a small incommensurate moment is induced outside the underdoped disk.
The LDOS along the line from R to B is shown in
Fig.~\protect\ref{fig_3}.}
\label{fig_1}
\end{figure}

Figure \ref{fig_1} shows typical results for an underdoped
superconductor for the self-consistent hole density $n_{hi} = 1-n_i$
with $n_i =\langle \hat n_i \rangle$, the staggered AF moment $m_i =
\frac 12 e^{i\Q\cdot \rr_i}(n_{i\uparrow}-n_{i\downarrow})$ with $\Q =
(\pi,\pi)$, and the $d$-wave order parameter $\Delta^{(d)}_i = \frac
12 \sum_\delta (-1)^{\delta_y} \Delta_{i\, i+\delta}$ where $\delta$
is summed over the four nearest-neighbor sites.  The average hole
density is $n_h\equiv 1-n \approx 0.07$ holes/site, but the hole
distribution is quite inhomogeneous.  In a homogeneously doped sample
with $J=1.5$ and $U=3.2$ (these are typical for this work), there is a
first order phase transition between SC and AF phases at a hole doping
level $n_h= 0.07$.  In the inhomogeneous system, the situation is more
complicated.  The spin polarization saturates near its bulk value in
undoped regions whose diameter exceeds $\xi_{\mathrm{AF}} \sim t_1 /
Um$, where $m$ is the staggered moment.  In smaller underdoped
regions, the staggered moment is roughly proportional to the diameter
of the region.  It is worth stressing that this behavior is very
different from single-phase models in which the magnitude of the local
order parameter is directly correlated with the local charge
density,\cite{engelbrecht} regardless of the size of the domain.  Note
also that, although the doped and undoped regions in Fig.~\ref{fig_1}
lie firmly on either side of the first order phase transition
separating AF and SC phases, both order parameters are finite
throughout the system because of a pronounced proximity effect.  In
this sense, the introduction of disorder in the doping leads to a
qualitative change in the phase diagram.  This aspect of the
calculations appears to be consistent with neutron scattering studies
in LSCO\cite{lake} suggesting that AF and SC coexist locally.  One
factor which appears inconsistent with experiment is that the $d$-wave
order parameter is suppressed by static AF correlations, whereas there
is good evidence that it actually grows rapidly as the insulating
phase is approached in HTS.  This disparity may be the result of the
simplicity of the current mean-field approximation.  In a more
sophisticated treatment, the suppression of SC order will be
compensated to some extent by the fact that the pairing interaction is
doping dependent: Numerical studies of the $t$-$J$
model\cite{zhangrice,randeria} find that $J\sim (1+n_h)^{-2}$.  Since
this result was originally derived for homogeneous systems, and cannot
be trivially extended to inhomogeneous systems (but should not change
our conclusions qualitatively), I will defer its discussion rather
than introduce an {\em ad hoc} local renormalisation of $J$.

\begin{figure}
\includegraphics[width=\columnwidth]{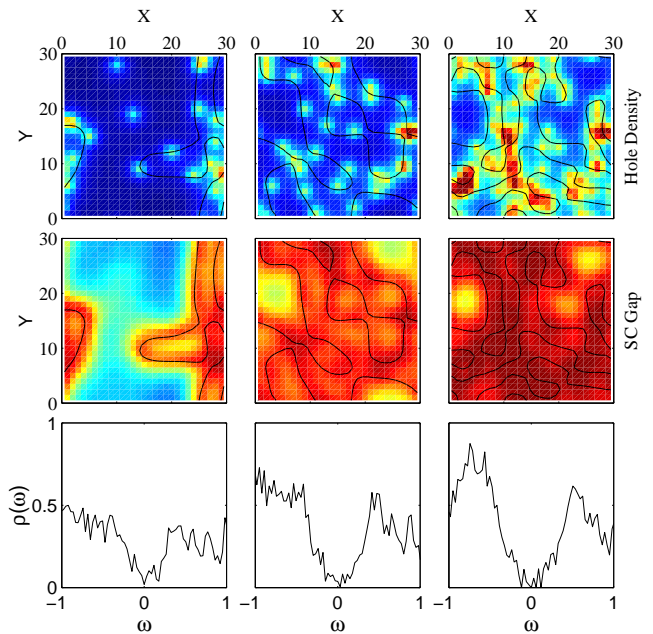}
\caption{Doping dependence of the AF and SC phases.  The three rows
display the hole density (top), SC gap (middle) and average density of
states (bottom) for three different donor-impurity concentrations.
The columns correspond to 20 donor-atoms (left) 35 donor-atoms
(middle) and 70 donor-atoms (right). Contours show the domain walls of
the staggered AF moments.  Parameters are as in
Fig.~\protect\ref{fig_1}. The color scales are identical for all
panels within a row, and are the same as in
Fig.~\protect\ref{fig_1}(a) and (c).}
\label{fig_2}
\end{figure}

Figure \ref{fig_2} shows how the coexisting phases evolve with hole
doping.  At low doping, the situation qualitatively resembles a pinned
smectic.  Smectic phases have been proposed as a natural mechanism by
which doped antiferromagnets can accomodate holes while minimizing
both the hole kinetic, antiferromagnetic exchange, and long range
Coulomb energies.\cite{kivelson} Consistent with the smectic picture,
the AF moments in Fig.~\ref{fig_2} spontaneously form $\pi$-shifted
domains whose boundaries are pinned to donor-impurity locations, and
the SC order parameter $\Delta^{(d)}_i$ is largest along the domain
walls.  However, there are a number of differences between the current
``weak coupling'' mean-field calculations and the canonical
``strong-coupling'' smectic picture.  First, because of frustration
introduced by next-nearest neighbor hopping, the AF phase is never
fully polarized.  This same frustration leads to a gapless
quasiparticle spectrum in the pure AF phase provided $U$ is less than
some model-dependent critical value.  As a consequence, holes are not
confined to domain walls but are mobile throughout the volume of the
sample.  In other words, although there are static AF correlations,
the system is on the metallic side of the metal-insulator transition.
A further consequence, which distinguishes weak and strong-coupling
approaches, is that the SC order parameter remains finite everywhere
in the weak-coupling calculations.

As hole doping increases, the AF phase is suppressed through a
proliferation of domain walls.
In Fig.~\ref{fig_2}, significant AF moments form at higher doping only
in regions where, due to randomness in the donor-atom distribution,
undoped regions have diameters larger than $\xi_{\mathrm{AF}}$.  The
system is then better understood as consisting of isolated AF pockets
embedded in a relatively homogeneous $d$-wave superconductor.  In this
model, annealing (which tends to homogenize the charge distribution)
will have significant effect on the extent of AF order.

Figure \ref{fig_2} also shows the spatially averaged density of states
$\rho(\omega)$ for each of the disorder configurations.  Although the
data is noisy, several clear features are evident: First, as one
underdopes, there is a gradual suppression of spectral weight on an
energy scale which is large relative to the SC gap.  Second, there is
a robust $d$-wave gap, even in situations where a large fraction of
the sample is antiferromagnetic.  I emphasize that this latter effect
is not necessarily anticipated since, in the absence of a nested Fermi
surface, AF ordering tends to destroy the particle-hole symmetry of
the spectrum.  Third, as one underdopes, the superconducting coherence
peaks are suppressed.  As discussed in the introduction, the
suppression of coherence peaks is one of the hallmarks of the
pseudogap phase of the underdoped cuprates.  To my knowledge, this is
the first reproduction of such an effect in terms of a static mean
field model.

These results are rather encouraging and, ideally, the next step should be
a detailed examination of the LDOS.
However, finite size effects limit the spectral resolution
to the extent that the LDOS is impossible to interpret.  
Consequently, I will focus  for the remainder of the paper
on intermediate doping levels where one can improve the energy
resolution by studying isolated underdoped pockets embedded in a large
superconducting domain.  A more detailed exploration of the pinned
smectic phase requires a different approach and is, unfortunately, beyond
the scope of this work.

\subsection{Single underdoped pocket}
\label{sec2b}
\begin{figure}
\includegraphics[width=\columnwidth]{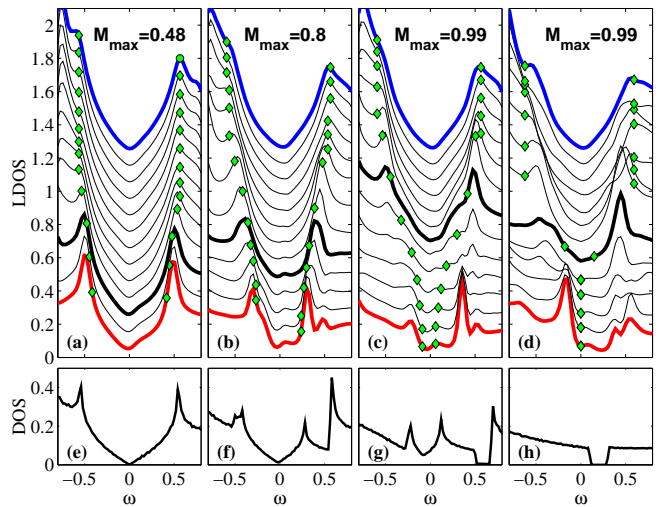}
\caption{Local density of states. (a)-(d) The LDOS is shown for
different model parameters.  The spectra, offset for clarity, are
taken at a sequence of sites extending radially outwards along the
$(010)$ direction from the center of an isolated underdoped disk of
radius $R$.  The path along which spectra are measured is shown in
Fig.~\protect\ref{fig_1}(d), with points B and R corresponding to the
blue (top) and red (bottom) curves in this figure.  The heavy black
curve in each panel indicates the site at which the staggered
magnetization falls to half the maximum value $m_{\mathrm{max}}$.
Diamond symbols indicate $\pm2\Delta^{(d)}_i$ (the estimated coherence
peak energies) at each site. The first three panels are for (a)
$U=3.2$, $R=1.5a_0$, (b) $U=3.2$, $R=4.0a_0$, and (c) $U=3.4$,
$R=4.0a_0$.  $J=1.5$ throughout.  In (d), spectra are calculated for a
non-self-consistent model of a pure antiferromagnetic pocket of radius
$R=6.0$ with $U=3.4$ embedded in a pure $d$-wave superconductor.  The
heavy black curve marks the sharp boundary between the AF and SC
domains.  For comparison, the DOS for homogeneously coexisting AF and
SC order are shown in (e)-(h).  Cases are (e)
$(M,\Delta^{(d)})=(0,0.3)$, (f) $(0.3,0.3)$, (g) $(0.6,0.3)$, (h)
$(1.0,0)$.}
\label{fig_3}
\end{figure}

It is difficult to discuss the STM spectrum in detail for finite-sized
lattices because of the discreteness of the spectrum.  For a $30\times
30$ lattice, one might typically have $\approx 100$ subgap states,
with the resulting spectrum being too noisy for anything other than
the grossest analysis.  I therefore study a single, isolated, AF
pocket which is embedded in a homogeneous background potential
corresponding to a hole doping level of $p\approx 0.15$ (in fact, the
hole doping level is less important than the Fermi surface shape, and
should not be taken too seriously).  In this calculation, a positively
charged disk of radius $R$ sits $d_z=1.5a_0$ above the conducting
layer.  The charge on the disk is adjusted so that the site under the
center is half-filled.  The charge, magnetization and SC gap are
calculated self-consistently, and an example of the self-consistent
magnetization for a disk of radius $R=4a_0$ is shown in
Fig.~\ref{fig_1}(d).  In order to obtain a high spectral resolution,
the underdoped pocket is embedded in a homogeneous $200\times 200$
region and a recursion technique\cite{haydock} is used to calculate
$\rho(\rr,\omega)$.  In this way, I avoid spurious structures
associated with the discreteness of the spectrum on finite
lattices. Figure~\ref{fig_3} shows the LDOS along cuts through the
centre of an AF pocket for different values of $\{ U,R \}$.

The spectra in Fig.~\ref{fig_3} show a smooth evolution from regions
where the $d$-wave order parameter is dominant, to the central region
where AF correlations are large.  The particle-hole asymmetry at large
energies comes from a van Hove singularity at $\omega \approx
-0.5|t_1|$.  At lower energies, the spectrum is determined by the
interplay between SC and AF order. There are several noteworthy
features of this calculation. First, there is an overall suppression
of spectral weight at site $i$ on an energy scale $M_i = Um_i$.  For a
nested Fermi surface, with homogeneous magnetization, $M_i$ is the
energy of the AF gap, but in the absence of nesting, AF correlations
lead to a shift of states away from the Fermi level even when a true
gap does not open.  This shift is a precursor to the formation of
lower and upper Hubbard bands, and is consistent with the observation
of spectral weight shifts on a large energy scale as a function of
doping in the cuprates.  Two factors tend to suppress the nucleated
moment: the frustration introduced by the next-nearest neighbor
hopping (ie.\ the absence of complete nesting of the Fermi surfaces),
and the competition with the superconducting phase.  I want to
emphasize a consequence of this which may not be intuitive: Although a
large $U$ may not generate a substantial moment $m$, the energy scale
$M$ over which the quasiparticle spectrum is affected by magnetic
correlations can still be quite large.  This is likely to be a
universal feature of models of competing order in the cuprates.  One
could similarly imagine that, in a model with a charge ordered phase,
frustration due to imperfect Fermi surface nesting and competition
with SC order would tend to suppress the magnitude of charge
modulations but still affect the spectrum over a relatively large
energy scale.

A second feature of Fig.~\ref{fig_3} is that the local dispersion in
the AF pocket near the Fermi level is quasilinear when $M_i \lesssim
\Delta^{(d)}_i$ [Fig.~\ref{fig_3}(a)]. This is a significant result
since one of the main arguments against coexisting commensurate order
in the cuprates is the absence of a linear dispersion at the Fermi
level.  When $M_i \gtrsim \Delta^{(d)}$, the LDOS becomes
particle-hole asymmetric: as $M_i$ increases (either as one moves into
the AF pocket, or as one turns up $U$), a shoulder develops in the
dispersion at low energies, which ultimately evolves into a
well-defined resonance [Fig.~\ref{fig_3}(b)].  The energy of the
resonance depends on the details of the band structure, and on
$\Delta^{(d)}$, but is a universal feature in nearly all numerical
results.

One of the most interesting aspects of Fig.~\ref{fig_3} is the
evolution of the coherence peaks between the SC and AF regions.  There
are actually two qualitatively different ways in which this occurs.
In Figs.~\ref{fig_3}(a) and (b), the coherence peaks shift to lower
energies, sharpen, and lose spectral weight as one moves into the AF
domain.  The coherence peak positions approximately reflect the local
value of $\Delta^{(d)}$, which is what one might naively expect for a
smoothly varying Hamiltonian.  The situation is different in
Fig.~\ref{fig_3}(c) where the magnetization is larger: the coherence
peaks (starting at a point exterior to the AF domain and moving
inwards) rapidly collapse, but shift rather little.  The absence of a
shift indicates that one is seeing the decaying tails of bulk BCS-like
states.  In other words, antinodal quasiparticles from the SC domain
{\em tunnel} (rather than propagate freely) into the AF domain, and
decay over some characteristic distance which determines the extent of
the coherence peaks into the AF domain.  I will argue below that this
arises from a mismatch in the SC and AF energy dispersions: when $M$
is sufficiently large, the states at the antinodal $k$-vector are
gapped in the AF domain.

The fact that the coherence peaks sharpen as one moves into the AF
region in Fig.~\ref{fig_3}(a) and (b) is the result of the fact that
the Fermi surfaces nest at isolated points (the contours
$\epsilon_\kk=0$ and $\epsilon_{\kk+\Q}=0$ shown in Fig.~\ref{fig_4}
intersect at two points).  Because of this peculiar nesting, spectral
weight is removed at energies both above and below the antinodal
saddle point energy (the point which generates the coherence peaks) by
the AF correlations, but the saddle point itself survives until the AF
moment becomes very large.  Thus, the coherence peaks lose weight by
narrowing rather than by being suppressed.  Other competing phases
(such as charge density waves) which nest differently should have a
qualitatively different effect on the coherence peaks.

One surprising aspect of Fig.~\ref{fig_3}(b) and (c) is that although
the transition from SC to AF domains occurs differently depending on
the magnetization, the spectrum at the core of the AF pocket is quite
similar. For comparison, a non-self-consistent calculation is shown in
Fig.~\ref{fig_3}(d) for the ansatz
\begin{eqnarray*}
\Delta_{ij}&=&\left\{ \begin{array}{cc} 
  0, & |\rr_i|<R \mbox{ or } |\rr_j| < R \\
  0.3, & \mathrm{otherwise}
  \end{array}
\right . , \\
m_{i}&=&\left\{ \begin{array}{cc} 
  0.3, & |\rr_i|<R \\
  0, & \mathrm{otherwise}
  \end{array}
\right . ,
\end{eqnarray*}
with $R=6a_0$.  Again, the spectrum at the core of the AF disk is
quite similar to that of the self-consistent calculations shown in
Fig.~\ref{fig_3}(b) and (c). (note the similarity in the peak
positions), but bears little resemblance to the spectrum of the
macroscopic AF phase shown in Fig.~\ref{fig_3}(h).  This calculation
demonstrates that the boundary conditions (the coupling between the AF
pocket and the SC bulk) have as large an impact on the local spectrum
at the core of the AF pocket as the local value of the SC order
parameter.  This is a central result: when the scale of the
inhomogeneity is atomic, one cannot assume a direct correspondence
between the local ordering and the local spectrum.  In the following
section, I argue that a qualitative understanding of the
inhomogeneously doped system can be developed from a model of {\em
homogeneously} coexisting SC and AF order.


\subsection{Homogeneously coexisting order}
\label{sec2c}
I consider a three-band model of a homogeneous system with coexisting SC and
AF long range order.  The SC order parameter has the usual form
$\Delta_\kk = \Delta^{(d)}(\cos k_x - \cos k_y)$ and the AF order
parameter is $M$ with $m_i =  M/U$.  I adopt the same
dispersion as before, with $\epsilon_\kk = t_0 +2t_1(\cos k_x + \cos
k_y) + 4t_2\cos k_x \cos k_y + 2t_3(\cos 2k_x + \cos 2k_y)$ and $t_0 = 0.7|t_1|$.  For a
complex frequency $z=\omega+i0^+$, the Green's functions satisfy
\begin{equation}
\left [ \begin{array}{ccc}
z -  \epsilon_{\kk} & \sigma M & \sigma \Delta_\kk \\
\sigma M & z- \epsilon_{\tilde \kk} & 0 \\
\sigma \Delta_\kk & 0 & z+\epsilon_{-\kk} 
\end{array} \right ]
\left [ \begin{array}{c}
G^{\sigma}_{\kk\kk}(z) \\
G^{\sigma}_{\kk\, \tilde{\kk}}(z) \\
\overline{F}^\sigma_{\kk\kk}(z) 
\end{array} \right ]
=
\left [ \begin{array}{c} 1 \\ 0 \\ 0 \end{array} \right ],
\label{bulk}
\end{equation}
where 
$\tilde{\kk} = \kk-\Q$, and where
$G^\sigma_{\kk\kk^\prime}(\omega) $ and
$\overline{F}^\sigma_{\kk\kk^\prime}(\omega)$ are Fourier transforms
of the retarded and anomalous Green's functions:
\begin{eqnarray*}
G^\sigma_{\kk\kk^\prime}(t) &=& -i\langle \{c_{\kk\sigma}(t),
c_{\kk^\prime\sigma}^\dagger (0)\} \rangle \Theta(t)\\
\overline{F}^\sigma_{\kk\kk^\prime}(t) &=& -i\langle \{
c^\dagger_{-\kk\, \sigmab}(t), c^\dagger_{\kk^\prime \sigma}(0) \}
\rangle \Theta(t).
\end{eqnarray*}
\begin{figure}
\includegraphics[width=\columnwidth]{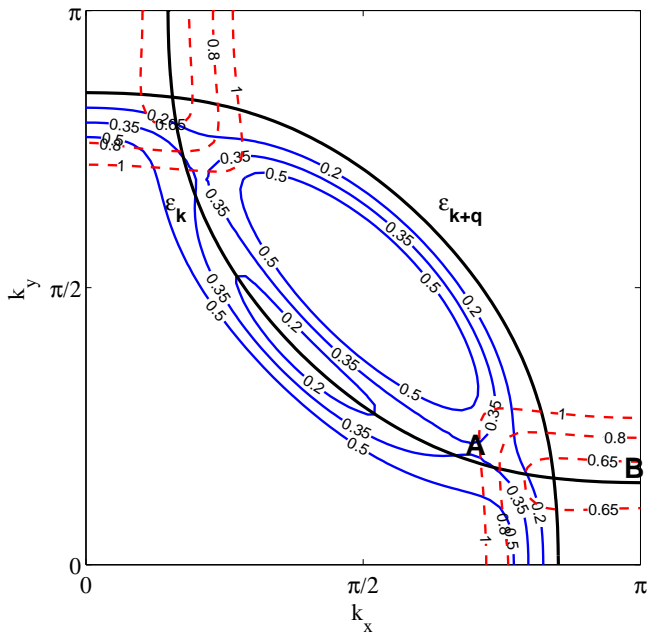}
\caption{Constant energy contours of the coexisting AF/SC model.
Parameters are the same as Fig.~\ref{fig_3}(e). A few contours are
shown for positive energies for the upper (dashed red) and middle
(solid blue) bands.  The zero-energy contours of $\epsilon_\kk$ and
$\epsilon_{\kk+\qq}$ are shown for reference. For small energies, the
contours have the ``banana''-like shape expected for $d$-wave
superconductors, as well as an antiferromagnetic shadow band. There is
a saddle-point singularity at $\omega=0.38$ (labelled ``A'' in the
figure) which marks the end of the linear dispersion, and gives rise
to the $\omega=0.38$ van Hove singularity in Fig.~\ref{fig_3}(f).
There is a second saddle-point at B in the figure, which gives rise to
the superconducting coherence peaks. The energies of the saddle-points
depend on both $\Delta^{(d)}$ and $M$.  For sufficiently large $M$,
the saddle-point at B evolves into a simple band minimum,
corresponding to the lower edge of the upper Hubbard band.}
\label{fig_4}
\end{figure}
The density of states $\rho(\omega) = -\frac{1}{\pi} \text{Im }
\sum_{\kk,\sigma} G^\sigma_{\kk\kk}(\omega+i0^+)$, plotted in
Fig.~\ref{fig_3}(e)-(h), is determined by the poles of $G_{\kk
\kk}^\sigma(\omega)$.  For reference, a few constant-energy contours
of the spectrum are shown in Fig.~\ref{fig_4} for a case with
$M=\Delta^{(d)}=0.3$.  The interested reader is directed to
Ref.~[\onlinecite{chubukov}] for an extensive discussion of the normal
state spectrum of this model.  In the coexisting state, there are two
features of interest: first, at low energies the spectrum resembles
that of the pure superconductor and, second, there is a new
saddle-point singularity (at ``A'' in Fig.~\ref{fig_4}) which arises
because of the coexisting order.  In the limit $M \gg \Delta$, the
origin of the saddle-point is fairly transparent:
$G_{\kk\kk}^\sigma(z)$ has three poles corresponding to upper and
lower magnetic bands with dispersion $E_\pm =
(\epsilon_\kk+\epsilon_{\tilde\kk})/2 \pm [
(\epsilon_\kk-\epsilon_{\tilde\kk})^2/4 + M^2]^{1/2}$ and a hole-band
with dispersion $E_0 = -\epsilon_\kk$.  When $\Delta_\kk$ is nonzero,
there is an avoided crossing of $E_0$ and $E_-$ which results in the
saddle point.  Both features of the dispersion are evident in
$\rho(\omega)$ [Fig.~\ref{fig_3}(f)], which resembles the pure
$d$-wave superconductor at low $\omega$, and has a resonance at the
saddle-point energy, $\omega=0.38$.  This model
appears to capture several aspects of the inhomogeneous spectra in
Fig.~\ref{fig_3}(a)-(d). First, it predicts the overall suppression of
spectral weight on magnetic energy scales.  Second, it predicts the
occurence of a subgap resonance.  Third, it shows that while a
$d$-wave-like tunneling gap survives to fairly large values of $M$,
there is an inward shift in the position of the apparent ``coherence
peaks" as $M$ increases.  This is purely a band structure effect
resulting from the reduction of spectral weight at the antinodal
points, and has nothing to do with a reduction of $\Delta^{(d)}$.
Fourth, it suggests that because AF nesting does little to disrupt the
band structure near the nodal points, there will be a much smaller
Fermi surface discontinuity for nodal quasiparticles crossing between
SC and AF domains than for antinodal quasiparticles.  This mechanism
is one possible explanation for relative uniformity of the low energy
spectrum measured in STM experiments when compared with the spectrum
near the gap edge.

This model makes one other, somewhat subtle, prediction which appears
to be relevant to the numerical work.  Fig.~\ref{fig_4} shows several
constant energy contours for the middle and upper bands.  For small
$M$, the bottom of the upper band lies below the top of the middle
band.  As $M$ is increased, however, a gap will appear in
$\rho(\omega)$.  Because the nesting points (the points at which
$\epsilon_\kk =\epsilon_{\kk+\Q}$) lie near the antinodal points, the
gap, when it opens, does so near the coherence peak energy, as in
Fig.~\ref{fig_3}(g). [Notice that the energy at which the gap appears
depends on both the band structure, and on $\Delta^{(d)}$.  Hence, the
gap in Fig.~\ref{fig_3}(g) appears at a higher energy than for a pure
antiferromagnet.  Furthermore, $\Delta^{(d)}$ tends to enhance the
magnitude of the gap.]  The key difference between Fig.~\ref{fig_3}
(b) and (c), where the coherence peaks evolve smoothly in the former
and collapse in the latter, appears to be the presence of an AF gap in
the spectrum.  I remark that while there is a threshold value of $M$
at which a gap forms, other kinds of order may not have such a
threshold.  For example, a charge density wave which nests between
parallel antinodal sections of the Fermi surface may gap out the
coherence peaks for any degree of ordering.

A second consequence of having a gap in $\rho(\omega)$ is that
scattering resonances may produce exponentially localized bound states
at energies within the energy gap.  The resonance at $\omega=0.35$ in
Fig.~\ref{fig_3}(c), for example, is very sharp, and localized to only
the few sites nearest the core of the AF domain.  Furthermore, it's
energy is close to where the spectral gap opens in the pure AF, making
it a good candidate for the kind of local resonance discussed here.
There are several sources of scattering---the inhomogeneity of the SC
and AF order parameters, and the impurity Coulomb potential---which
are not included in the homogenenous model, which could give rise to a
bound state.

\subsection{Weak charge modulations}
\label{sec2d}
\begin{figure}
\includegraphics[width=\columnwidth]{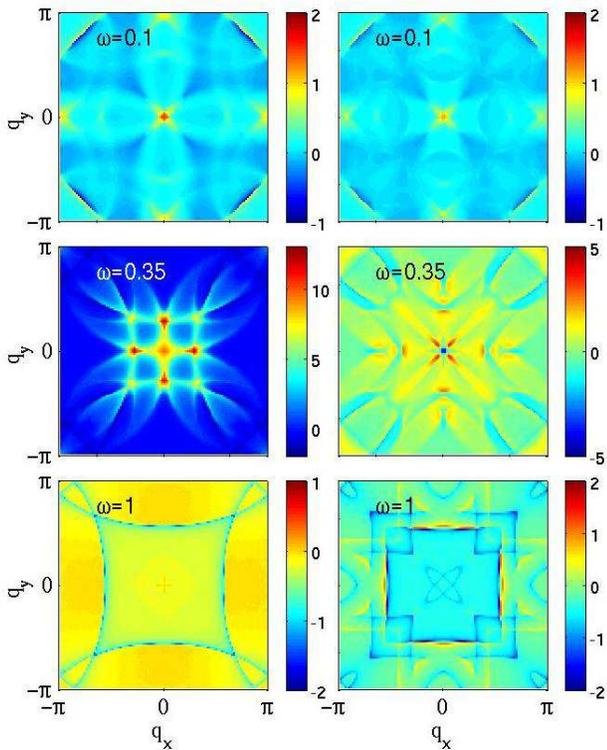}
\caption{Response kernel for disorder. The imaginary part of
$\Lambda_3(\qq,\omega)$ is shown for three values of $\omega$ for the
coexisting order (left column) and pure superconducting (right column)
cases.  The parameters are the same as in Fig.~\protect\ref{fig_3}(f)
(left column) and (e) (right column).  At low energy, the kernel is
similar for both models (top row).  The saddle point (labelled ``A''
in Fig.~\protect\ref{fig_4}) causes the peak at $\qq\approx (\pi/2,0)$
to split into two at $\omega=0.35$.}
\label{fig_5}
\end{figure}

Weak charge modulations have been observed in underdoped
BSCCO,\cite{Kapitulnik3,McElroy} sparking an ongoing debate as to the
extent to which spatial modulations of the LDOS can distinguish
Friedel oscillations of quasiparticles from a tendency towards
charge-ordering.  In this section I will address two slightly
different questions which arise from this debate.  First, I will
discuss charge ordering in the context of the current calculations.
Second, I will discuss the broader question of the extent to which a
hidden order (an order which does not couple directly to the charge)
can be revealed by quasiparticle scattering.

In Sec.~\ref{sec2a}, the self-consistent calculations show that a
large charge inhomogeneity occurs in the hole-doped domains, and
arises because of randomness in the impurity locations.  In contrast,
the undoped domains are remarkably homogeneous simply because they are
free from impurities.  The charge inhomogeneities are not particularly
evident at energies near the Fermi level, but become apparent at large
energies.  The calculations do not find a local charge ordering,
although the kind of weak modulations seen in
Ref.~[\onlinecite{McElroy2}] would be difficult to see on the
finite-size lattices used in numerical calculations.  Even if the
current calculations do not admit charge ordering, it is likely that
only minor modifications to Eq.~(\ref{HU}) are needed to generate
charge-ordered phases.  By analogy with the self-consistent solutions
for the AF phase, one would expect that these charge ordered phases
would coexist with superconductivity throughout the system because of
the proximity effect.  This is in contrast to experiments in
BSCCO,\cite{McElroy2} however, which see ordered charge modulations
only in ``pseudogap" domains.  A resolution to this puzzle may be the
fact that the charge distribution is uniform in underdoped domains,
but is disordered in hole-doped domains.  If the CDW order parameter
is easily pinned by donor atoms, then it may also be locally
suppressed by donor-related disorder.

The second question is whether hidden order can be revealed through
the Fourier transformed density of states of a disordered superconductor.
In order to get some sense of how disorder affects the LDOS, I return to 
the model of
homogeneously coexisting order described in the previous section.  I
calculate the response kernel,\cite{capriotti,ZAH} 
\begin{eqnarray*}
\Lambda_3(\qq,\omega) &=& \sum_\kk
\mathrm{Tr_\sigma } [G^\sigma_{\kk \kk}(\omega) 
G^\sigma_{\kk+\qq\, \kk+\qq}(\omega) \\
&&- F^\sigma_{\kk\kk}(\omega)
F^\sigma_{\kk+\qq\,\kk+\qq}(\omega)],
\end{eqnarray*}
which describes the effects of scattering from impurities on the
Fourier transformed density of states $\rho(\qq,\omega)$.  The results
are shown in Fig.~\ref{fig_5}.  At low $\omega$,
$\Lambda_3(\qq,\omega)$ is similar for both coexisting order and for
the pure $d$-wave superconductor while, at higher energies, the
effects of antiferromagnetism become significant.  At $\omega=0.35$
(the energy of the saddle point marked ``A'' in Fig.~\ref{fig_4}) the
pronounced resonances along the $(\pi,0)$ and $(0,\pi)$ directions are
split by the antiferromagnetism.  One surprising result of these
calculations is that $\Lambda_3(\qq,\omega)$ differs significantly
from the pure $d$-wave result even for energies much larger than $M$,
suggesting that the effects of even weak ordering should be easily
visible.  At the same time, however, there is no obvious signature of
the AF $\Q$-vector in the response kernel at most energies.  In other
words, AF ordering distorts the response kernel from the bare kernel,
but does so in a nontrivial way.  In particular, all features in
$\Lambda_3(\qq,\omega)$ disperse with $\omega$.  Thus, it appears that
unless the nucleated order couples directly to the charge density, it
will generally be difficult to distinguish different kinds of order
from the Fourier transformed density of states.

\section{Conclusions}
\label{sec3}
I have studied a mean-field extended Hubbard model in which charge is
doped inhomogeneously because of randomness in the donor-atom
positions.  AF and SC order compete, and in the homogeneous case, are
separated by a first order phase transition.  Self-consistent
calculations find that, because of inhomogeneity in the local doping,
AF order coexists inhomogeneously with superconductivity.  The AF
moments spontaneously form $\pi$-shifted domain walls which are pinned
to donor-atom sites.  At low doping, the self-consistent solutions
resemble pinned smectics, ie.\ quasi-one-dimensional superconductors
running along AF domain walls.  Because of the proximity effect, both
SC and AF correlations are actually present throughout the lattice.
This picture appears to be consistent with neutron scattering studies
in LSCO,\cite{lake} suggesting that AF and SC order coexist, and
naturally explains the field-dependence of the AF moment, since any
suppression of superconductivity by a magnetic field will enhance the
AF moment.

At higher doping, the self-consistent calculations evolve towards a
homogeneous $d$-wave superconductor interspersed with underdoped
pockets with large AF moments.  This latter ``phase-separated'' system
superficially resembles the situation in BSCCO, although it is
generally unclear whether the ``pseudogap'' domains in BSCCO have any
kind of nucleated secondary phase.  To address this question, I
studied the local spectrum of a single, isolated AF pocket embedded in
a homogeneous SC background.  While no single calculated spectrum
reproduces all details of the experimental measurements, several
features such as the collapse of the coherence peaks, the occurence of
low-energy spectral features, and the relative homogeneity of the low
energy spectrum, are broadly consistent with the kinds of spectra
measured in, for example, Ref.\ [\onlinecite{McElroy2}].  Certain
experimental aspects---notably the presence of weak nondispersing
charge modulations---are not reproduced in my calculations.  In
general, the calculated spectra at low energies show a richer spectrum
of peaks than is observed experimentally.  

At this point, the effect of disorder on the spectrum of the isolated
AF pocket is not understood.  Earlier studies of point-like defects in
$d$-wave superconductors show that this may not be a trivial effect.
A single strong-scattering point-like impurity introduces a sharp
resonance near the Fermi level.  As the disorder level increases, the
resonances split, are inhomogeneously broadened, and evolve into an
impurity band (see Ref.~[\onlinecite{atkinsonI}] for a recent
summary).  When the response of the SC order parameter to the disorder
is included self-consistently, the SC gap tends to restore
itself\cite{atkinsonII} by shifting spectral weight away from the
Fermi level.  Indeed, it is a general feature of interacting electrons
in disordered media that the system can lower its energy by
suppressing the density of states at the Fermi level.  Further
calculations, currently in progress, are needed to establish 
whether all the spectral features discussed in Sec.~\ref{sec2b} survive
in the disordered limit.

Finally, although the calculations were performed for a model in which
superconductivity and antiferromagnetism compete, I expect many of the
findings to apply to other models of competing order.  Three
results, in particular, are expected to be general. First, when
the domain sizes are small (as they appear to be in BSCCO), the
proximity effect is extremely important, and has a significant impact
on the local density of states.  It was {\em never} found, in
calculations, that the spectrum of the antiferromagnetic pocket
resembles that of a bulk antiferromagnet.  Rather, a better toy model
appears to be one of coexisting homogeneous superconductivity and
antiferromagnetism.  Second, the gapping of the spectrum near the
antinodal points by local ordering is a mean-field mechanism by which
coherence peaks may be locally suppressed.  Up to now, it has been
generally understood that suppression of coherence peaks occurs
through strong inelastic scattering at higher energies.  For AF order,
there is a threshold value of the magnetization for the gapping of the
antinodal quasiparticles, but this may not be universal and other
kinds of order (eg.~CDW) may lead to suppression of coherence peaks
for even small ordering.  Finally, these calculations also suggest a
natural reason that nodal quasiparticles should be less affected than
antinodal quasiparticles by charge inhomogneities, since the Fermi
surface mismatch between domains is smallest for the nodal
quasiparticles.

\section*{Acknowledgments}

The author would like to acknowledge helpful conversations with
P. J. Hirschfeld and R. J. Gooding.  This work was supported in part
by Research Corporation grant CC5543.

\end{document}